\documentclass[10pt]{article}
\usepackage[OE]{express}
\usepackage[nomargin,inline,draft]{fixme}
\fxusetheme{color}
\begin{document}

\title{Neuromorphic photonics with electro-absorption modulators}

\author{Jonathan George,\authormark{1} Armin Mehrabian,\authormark{1} Rubab Amin,\authormark{1} Jiawei Meng,\authormark{1} Thomas Ferreira de Lima,\authormark{2} Alexander N. Tait,\authormark{2} Bhavin J. Shastri,\authormark{2} Tarek El-Ghazawi,\authormark{1} Paul R. Prucnal,\authormark{2} and Volker J. Sorger\authormark{1*}}

\address{\authormark{1}Department of Electrical and Computer Engineering, George Washington University, Washington DC, 20052, USA\\
\authormark{2}Department of Electrical Engineering, Princeton University,  Princeton NJ, 08544, USA\\}

\email{\authormark{*}sorger@gwu.edu} 



\begin{abstract}
Photonic neural networks benefit from both the high channel capacity- and the wave nature of light acting as an effective weighting mechanism through linear optics. The neuron's activation function, however, requires nonlinearity which can be achieved either through nonlinear optics or electro-optics. Nonlinear optics, while potentially faster, is challenging at low optical power. With electro-optics, a photodiode integrating the weighted products of a photonic perceptron can be paired directly to a modulator, which creates a nonlinear transfer function for efficient operating. Here we model the activation functions of five types of electro-absorption modulators, analyze their individual performance over varying performance parameters, and simulate their combined effect on the inference of the neural network. 
%
\end{abstract}



\bibliographystyle{unsrt}
\bibliography{references}

\section{Introduction}
\begin{figure}[ht!]
\centering\includegraphics[width=13cm]{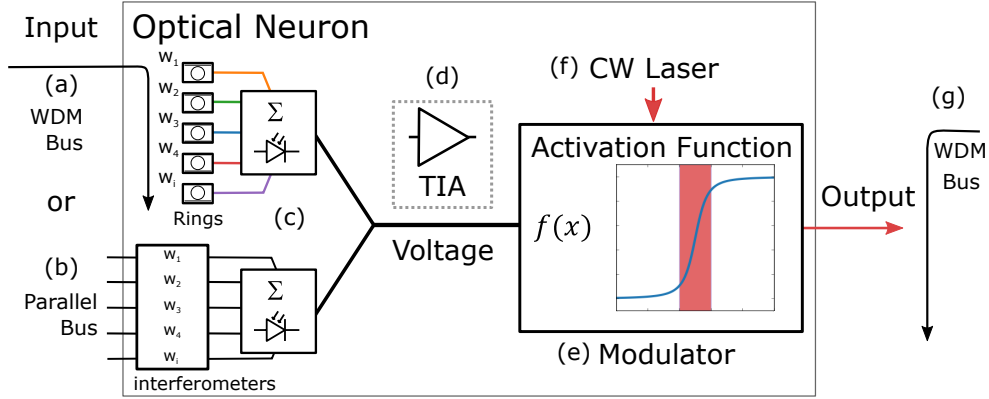}
\caption{An electro-optic neuron taking an input from a WDM bus and weighting by wavelength with rings (a) or from a parallel bus weighting with an interferometer network (b), sums the optical signal with a photodiode converting the signal to a voltage (c), optionally amplified by TIA (d), drives an electro-optic modulator (e) modulating a CW laser (f), produces a nonlinear transfer function at the output (g).\label{fig:overview}}
\end{figure}

Photonic neural networks (NN) have the potential for both high channel capacity (i.e. data baud rate) 
and low operating power. The former is provided via a) charing of small electrical capacitors and b) 'bosonification' where many photons are allowed to occupy the same quantum state, such as technologically utilized in wavelength division multiplexing (WDM). If the light-matter-interaction is enhanced such as realized via sub-wavelength photonics or plasmonics \cite{aJEOM,EOswitch,EOMmat}, the energy-per-compute (e.g. bit, or multiply-accumulate, MAC) can surpass electronic efficiency (i.e. about 1-10 GMAC/J) \cite{PNNperform} 
An artificial neuron requires two functions; first, it must have a synaptic-like linear function, weighting the set of inputs from other neurons (Fig. \ref{fig:overview}(a,b)). Secondly, it must apply a nonlinear activation function to the sum of the weighted inputs (Fig. \ref{fig:overview}(e)). In photonic neural networks, these functions can also be separated into a passive weighted interconnect and a set of active photonic neurons.
%
%
%

Interferometric \cite{shen2017deep} and ring-based \cite{tait2014broadcast} weighting have both been previously realized in integrated photonic platforms. With interferometric weighting, the phase of coherent light is utilized in a mesh of Mach-Zehnder interferometers to produce a vector dot product of the neuron's input vector and its weights. Similarly, in ring-based weighting, photonic rings are selectively tuned to apply a dot product, a potentially more compact method with wavelength-division multiplexing (WDM). In both cases linear optics achieves an efficient weighting where the wave nature of light computes the inner product simply by propagating forward in time. After the weights are tuned, the only additional energy consumed in creating the inner product is due to the additional laser power required to counteract propagation losses.
%


%
%
%

An activation function is a nonlinear function that is applied to the weighted sum of the inputs of a neuron (Fig. \ref{fig:overview}(e)). The nonlinearity of the activation function allows the network to converge into definitive states by eliminating infinitely cascading noise, similar to the nonlinearity of the transistor forcing the digital computer into binary states. Apart from the requirement of nonlinearity, and differentiability for training, there are no limits to the shape of the activation function itself and many activation functions have been proposed each with strengths in different applications \cite{kawaguchi2016deep}.
%
%
%
%
%
%

In the electro-optic neuron \cite{tait2017neuromorphic} considered here consisting of a photodiode connected either directly, or through an interface circuit, to an optical modulator, the nonlinear response of the modulator itself can be used to generate the necessary nonlinearity in the signal transfer function. In this case, the choice of modulator type will immediately impact the shape of the activation function and thus the operation of the network. Therefore the traditional design goals for optical modulators, while still relevant, must be evaluated with respect to cascadability and overall network performance.

In this paper, we first introduce the electro-optic absorption modulator and examine types of interface circuits for coupling absorption modulator based electro-optic neurons. Next, we introduce a method of clock-gated capacitive coupling, allowing a reduction in circuit resistance in capacitive modulator based electro-optic neurons. Next, we build a model of the capacitively coupled electro-optic neuron for four types of absorption modulators and compare their cascaded signal-to-noise ratios (SNRs) across modulator types with modulator length and laser power as swept parameters. Lastly, we evaluate the system performance of the four modulator types on the well-studied MNIST feeed-forward neural network using optimized parameters from the SNR analysis. Our results show the quantum well absorption modulator based electro-optic neuron outperforming the other three absorption modulators across a wide range of modulator lengths and optical powers.

%
%
%
%
%
%

\section{Modulators for Nonlinear Activation}
To set the context for the discussion of electro-optic modulators for NNs, in this work we are interested in photonic chip-based NNs, where integration density is a key value proposition. Modulators either alter the signals amplitude directly via electro-absorption (EAM), or shift the phase inside an interferometer (electro-optic modulator, EOM) such as in linear Mach Zehnder interferometers (MZI)  or microring resonators (MRR) to modulate the amplitude. To provide the nonlinear activation function in photonic NNs either modulator can utilized yet with different rationales; the advantage of EAMs is that they do not rely on interferometeric schemes to modulate a signals amplitude and, thus, can be designed more compact than EOMs. Reducing footprint allow increasing a) the neuron's areal density, and b) the photonic neurons' firing- or clock speed, since the MAC rate (i.e. MAC/s) scales with modulator 3dB-bandwidth (speed). However, if the VMM is performed shifting phase, then using phase for the activation function may be synergistic to the design layout. An advantage of field-driven EOM over carrier-based EAMs is, that their intrinsic switching effect is instantaneous as compared to electronic clocking \cite{Leuthold}. The details of EOM nonlinear activation functions will be reported elsewhere. We note, that both ref \cite{shen2017deep} and \cite{tait2014broadcast} used modulators only to perform the vector matrix multiplication and not the nonlinear activation. 

\begin{figure}[ht!]
\centering\includegraphics[width=13cm]{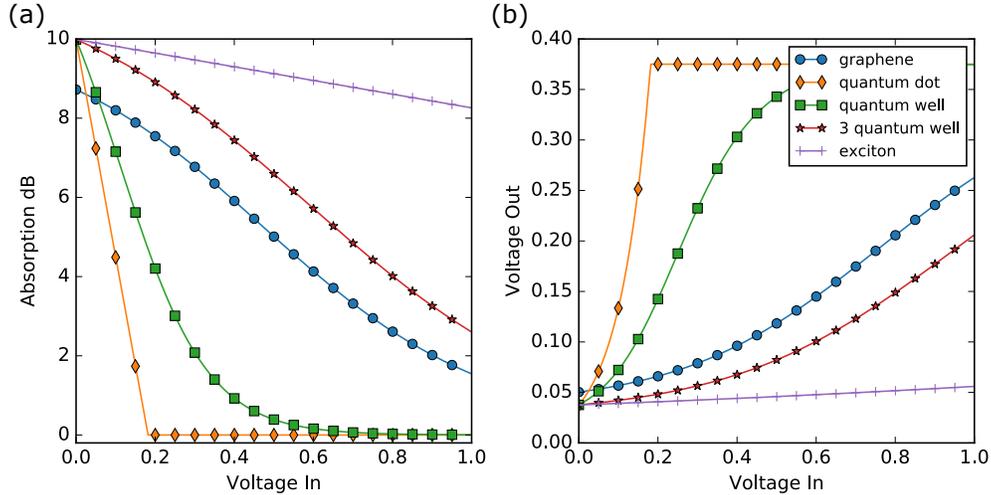}
\caption{Models of the nonlinear optical absorption of the electro-optic absorption modulators (EAM) vs. their drive voltage $ (V_{in}) $, equations derived in \cite{amin2017waveguide}, (a) varies by the type of tunable material used in the capacitively-biased modulator and translates into a voltage activation function $ (V_{out}) $ when a 100 $ \mu m $ modulator is coupled through a 50 $ \Omega $ current-to-voltage converting resistor to a photodiode (at the receiving neuron) with quantum efficiency of $\eta = 0.6$ (b). 
\label{fig:activationfunctions}}
\end{figure}

In this work we consider carrier-based EAMs only motivated by a) emerging material developments able of unity-strong optical index modulation which is 3-orders of magnitude stronger than the plasma effect in Silicon \cite{EOMmat}, and b) the chip density arguments made above. EAM devices absorb light as a function of their electrical bias. Either they absorb more light at zero bias and less light as the bias increases in magnitude or vice versa, depending on the type. This effect is used in optical communications to encode electrical signals on optical carriers. The shape of the voltage to absorption curve varies by EAM type \cite{amin2017waveguide} (Fig. \ref{fig:activationfunctions}). All EAM absorption curves are nonlinear due to the eventual saturation in the number of carriers (depletion or injection). 


\section{Photodiode Coupling}
The electro-optic mediated nonlinearity is created first by converting an optical signal into an electrical signal (O-E) and then by converting the electrical signal back into an optical signal (E-O). The performance of this nonlinear activation depends strongly on the choice of coupling between the photodiode and the electro-optic modulator performing the respective conversions. To this end, we model the photodiode electrically as a current source and the electro-optic modulator as a voltage dependent capacitive load. The current produced by the photodiode must then be converted into a voltage of sufficient magnitude to drive the modulator.

\begin{figure}[ht!]
\centering\includegraphics[width=10cm]{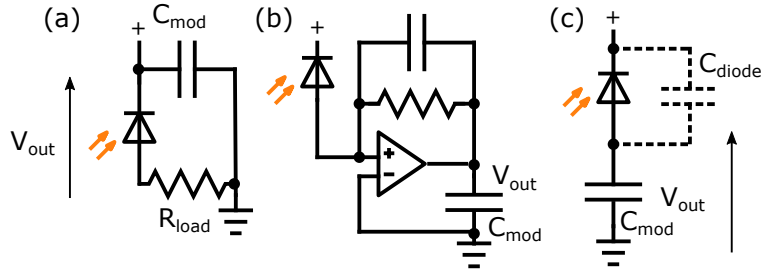}
\caption{The photodiode is coupled to the modulator with a reverse bias either either by a current to voltage converting resistive load, $ R_{load} $, (a), a TIA (b), or directly coupling to the modulator as a capacitive load (c). When coupled with load resistance $ R_{load} $ must be large enough to to produce a voltage in the operating range of the electro-absorption modulator, limiting the RC frequency response of the circuit.  \label{fig:photodiode_coupling}}
\end{figure}

There are four options for coupling the photodiode to the modulator; the first adds a current-to-voltage converting resistor to the circuit (Fig. \ref{fig:photodiode_coupling}(a)). This is the simplest method and can be used to produce a voltage of any magnitude. However, the resistor in combination with the capacitance of the modulator and photodiode acts as a low pass RC filter. The greater the required voltage, the greater the required resistance and the slower the device operates. The second method adds a transimpedance amplifier (TIA) to the circuit. This is the approach used by most optical receivers but is  costly in terms of circuit complexity, power, and noise figure. The third method is capacitive coupling, connecting the photodiode directly to the modulator \cite{miller2017attojoule} (Fig. \ref{fig:photodiode_coupling}(b)). Here the photodiode acts as a constant current source to charge the capacitive modulator. The final method is to inductively load the photodiode to transiently convert the low voltage of the photodiode to a higher voltage to drive the modulator. Like resistive coupling, this method creates an LC filter and requires an inductor of enough magnitude to create the necessary transient voltage, limiting operating speed.

Of these four coupling methods, capacitive coupling is the most appealing because it only requires scaling capacitance and not resistance or inductance both of which can be minimized, increasing the maximum operating frequency of the circuit, and hence of the entire photonic NN.

To model the capacitively coupled electro-optic circuit, we begin with the optical power of the CW laser source in Watts. The electro-optic absorption modulator will attenuate this signal with $ \alpha $ as a function of voltage.
\begin{equation}\label{power_dB}
P_{out} = P_{cw} exp \left(-\alpha\left(V_{in}\right)L\right)
\end{equation}
This optical power out of the modulator creates an arrival rate of photons $ \gamma_p $ at the next layer's photodiode. The event rate in the interval is modeled with a Poisson distribution, and is proportional to optical wavelength $ \lambda_0 $ and inversely proportional to operating frequency $ f $. The arriving photons act on the photodiode to move an elementary charge $ q $ with some quantum efficiency of $ \eta $, to charge a circuit with total capacitance $ C $, to ultimately reach a voltage change of $ \Delta V_{out} $ at the time $ 1/f $, assuming the reverse bias is much greater than $ \Delta V_{out} $. 
\begin{equation}\label{deltaV}
\Delta V_{out}=P_{poisson}\left(\gamma_p={\frac{\lambda_0}{fhc}P_{in}}\right)\frac{q\eta}{C} + V_{noise\_circuit}
\end{equation}
From Eq. \ref{deltaV}, we observe that adding an amplifier to the circuit, multiplying Eq. \ref{deltaV} by gain $ G $, is equivalent to increasing the input laser power, ignoring noise. By adding optical gain $ G $ to the laser, the voltage root mean square (RMS) noise is increased by

\begin{equation}\label{opticalNoiseGain}
\Delta N_{optical}=\sqrt{\gamma_p}\frac{q\eta}{C}\left(\sqrt G-1\right)
\end{equation}

To create equivalent gain with an electrical amplifier placed after the photodiode interfacing circuit, the increase in noise by the amplifier will be
%
%


\begin{equation}\label{amplifierNoise}
{\Delta N}_{amp\_gain}=G_{amp}\left(N_0\right)-N_0+N_{amp}, N_0 = \sqrt{\gamma_p}\frac{q\eta}{C}+N_{circuit}
\end{equation}

Where $ N_0 $ is the original noise of the circuit, now amplified by the new electrical gain G. $ N_0 $ is composed of the optical noise and the noise of the circuit, $ N_{circuit} $. It is interesting to note that adding an electrical amplifier, multiplies original optical noise, while adding optical gain does not.

Looking just at noise and ignoring power we would like to find when it is beneficial to add an amplifier to the circuit vs. increasing the operating optical power. This is when $ \Delta N_{amp\_gain} < \Delta N_{optical} $. Then, solving for amplifier noise, $ N_{amp} $, this is the region

\begin{equation}\label{amplifierRegion}
N_{amp\_gain}<\sqrt{\gamma_p}\frac{q\eta}{C}\left(\sqrt G+N_{circuit}\left(1-G\right)\right)
\end{equation}

The electrical power consumption to produce optical gain is increased by $ \left(G-1\right)P_0/\eta_{photodiode}\eta_{laser} $.

These two limits create a noise and power budget, respectively, in which an amplifier is beneficial to the circuit.

While capacitive coupling offers an efficient means of interfacing capacitive modulators to a photodiode, it also requires a clock to limit the capacitor charge time. If a clock is not used to gate the modulator's charge, the modulator will continue charging until the photodiode is no longer reverse biased. Furthermore, the layers must be activated in two cycles: in the first cycle a layer is charged, and in the following cycle it is held while the layer above it is charged. This effectively divides the throughput of the neural network in half.

There are two options available for implementing a clock. The first is to add an electrical gate to isolate the capacitor from the photodiode after it has been charged. This method adds additional power and design complexity to drive the electronic gate at each modulator. The second method is to replace the CW laser source feeding each layer of the neural network with pulsed sources. The pulses alternate between even and odd layers such that the lower layer is held while the upper layer is charged. This can be combined with an electronic gate to reset the layer at the beginning of the cycle or the modulator can be designed to leak charge at a set rate to reach a zero potential at the end of the held cycle.

\section{Noise and Cascadability}
In neuromorphic photonics, the neuron must exhibit both nonlinearity and a sufficient SNR at each layer to produce an SNR greater than one at the output of the final layer of the network. With a large number of layers, such as in deep-learning networks, the signal must cascade from layer to layer, and maintaining an SNR greater than one at the output of the network within a reasonable power budget. This requirement bounds the type of modulator, operating power, and number of achievable layers.

While the shape of the transfer function, including the nonlinearity, is primarily driven by the modulator type, the cascaded SNR of the electro-optic neuron is dependent upon several parameters. First, at the most rudimentary level, the input optical power generates a Poison distribution in the quantized arrival time of the photons. Next, the interfacing circuit adds thermal noise, and potentially gain. Finally, the electro-optic modulator itself affects the cascaded SNR through modulation depth and the modulator's intrinsic nonlinearity.

Immediately apparent in the analysis is the SNR's dependence on each node's operating power. The output SNR of the lowest power optical nodes, those nodes with the smallest input, will be less than the highest optical power nodes, those nodes with the greatest input. The cascaded SNR of the system, then, is dependent on the input data and the trained weights of the network.

It follows that in optimizing device parameters this forces us to make assumptions about the statistics of the input data and the trained weights. In our analysis of cascaded SNR we assume that the input power to each neuron during operation is uniformly distributed in the nonlinear portion of the voltage transfer function. We define this portion of the transfer function as the region with slope greater than 0.1 between the minimum and maximum swept voltage. We then evaluate the SNR of the cascaded network in terms of AC root mean square (RMS) power, Eq. \ref{ACpower}, for a uniformly distributed signal in this region.

\begin{equation}\label{ACpower}
P = \overline{\left(X^{2}-\bar{X}\right)^{1/2}}
\end{equation}

\begin{figure}[ht!]
\centering\includegraphics[width=13cm]{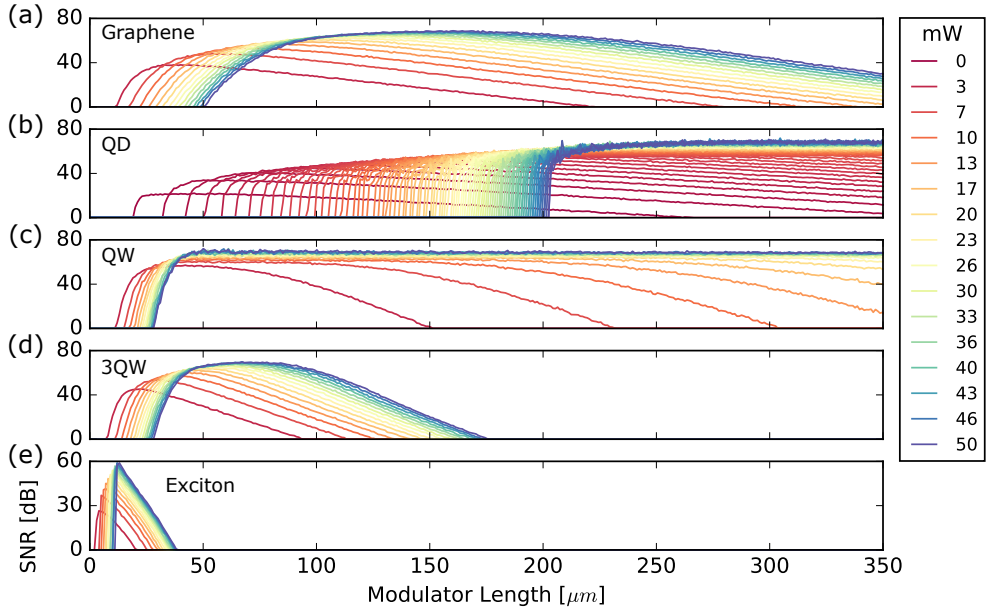}
\caption{
The AC SNR of a neurons output after two layers for graphene (a), quantum dot (b), quantum well (c), three quantum well (d), and exciton (e) plotted against modulator length over a range of optical powers from 0.01 mW to 50 mW shows low performance of graphene for low optical powers, almost no response for QD over for optical power over 10 mW, a wide range of reasonable performance for QW, and a short region of peak performance for exciton with modulator lengths < 350 $\mu m$. \label{fig:snr_1_mw_after_6layers}}
\end{figure}
\section{Training}
The electro-optic neuron produces an optical output from its weighted inputs. The power of this optical output must be distributed to the next layer's nodes. A naive layout would simply fan-out the output power to the next layer. This would create a 1/N power divider where N is the number of nodes in the next layer. In this scheme, the power to the next layer is severely limited. Even if the weights are trained to full power, 1, the full power from the divider is 1/N. Alternatively, ring drop filters in a broadcast and weight network or an MZI network can be used to divide the power among the output layers proportionally to the trained weights. In these configurable power dividing networks the total power ratio cannot exceed unity, i.e. no power is added. This limitation requires enforcing a constraint on the weight matrix during training to keep the sum of the output weights of any node from exceeding one. This is equivalent to enforcing a sub-unity L1 norm along the rows of a TensorFlow order (M Input x N Output) weighting matrix and is similar to the sub-stochastic matrices used in some Markov models \cite{pruitt1964eigenvalues}.

\section{Power Analysis}
The power dissipation of the capacitively coupled electro-optic modulator can be described by a charging and discharging capacitor. The energy stored in each circuit is defined by the energy of the capacitance, $E=CV^2/2$. Each node must charge the modulator in each operating cycle. 
 The total electric power dissipated ($P_e$) by the nodes of the network then is the charging and discharging of the capacitance of the modulator, photodiode, and gate at frequency $f$, $ P_e=N_{nodes}\left(V_N^2\left(C_{mod}+C_{photodiode}\right)+C_{gate}V_g^2\right)f/2$. In addition to the electrical power to operate the electro-optic modulator, the network requires a CW laser source of efficiency $\eta_{laser}$ with enough optical power to supply every node with enough optical power to reach the necessary SNR found in the previous analysis, $P_{laser}={N_{nodes}P}_o/\eta_{laser}$. The total power consumption of the network then becomes:
\begin{equation}\label{powerTotal}
P_{total}=N_{nodes}\left(\frac{\left(V_N^2\left(C_{mod}+C_{photodiode}\right)+C_{gate}V_g^2\right)f}{2}+\frac{P_o}{\eta_{laser}}\right)
\end{equation}

\section{Results and Discussion}
The MNIST dataset \cite{lecun-mnisthandwrittendigit-2010} is a set of images of handwritten digits in a grayscale 28x28 pixel format that is commonly used in comparing neural network performance. A network is trained to classify the images into the 10 individual digits and then evaluated on accuracy. We developed a simulation of a MNIST  classification neural network in Python using Keras \cite{chollet2015keras} and TensorFlow \cite{tensorflow2015-whitepaper}. 
The simulated optical neural network was derived from a standard MNIST design with two layers of 100 nodes per layer. The sigmoid activation function was replaced with a custom activation function implementing each of our derived electro-optical transfer functions, including noise. The weights were bound between zero and one to simulate input optical weighting by ring or interferometric modulators. The network was initialized with zero weights and trained with the Adagrad \cite{Duchi:EECS-2010-24} method, the categorical cross entropy loss function, a learning rate of 0.005, 45 training epochs, and no decay. An L1 less than unity constraint was placed on the rows of the TF ordered weight matrix during training to enforce power conservation in the optical weighting network. The simulated input laser power was swept in the range of 0.01 mW to 50 mW, and for each optical power the modulator length was selected by maximizing the AC SNR.

\begin{figure}[ht!]
	\centering\includegraphics[width=13cm]{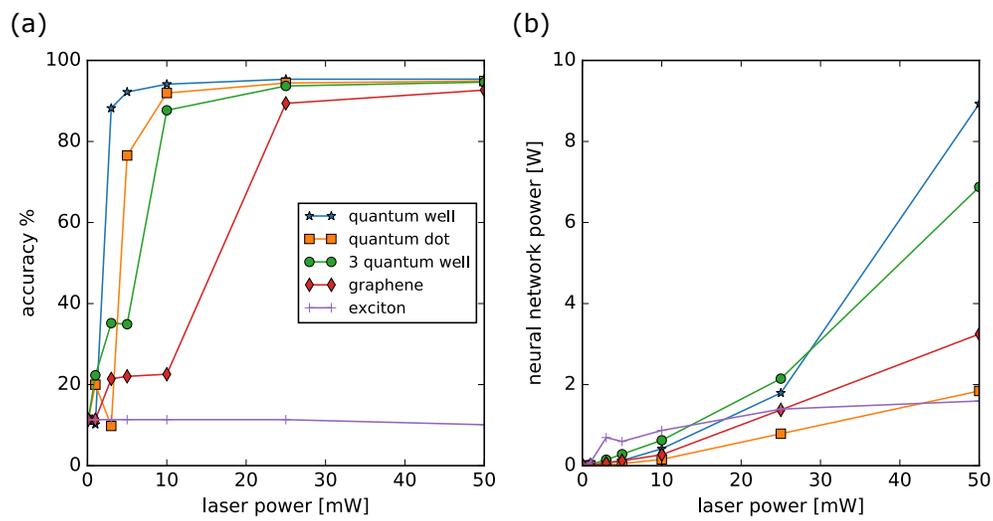}
	\caption{
		Simulation of a 200 node MNIST classification neural network over range of laser optical powers from  0.01 mW to 50 mW (a) show accuracy results converging across modulator types, except exciton, as optical power exceeds 30 mW. At lower power levels modulator types vary in performance with the quantum well modulator outperforms the others in terms of accuracy. Power dissipation, excluding electrical power to drive the CW laser and clocking overhead, (b) shows the quantum dot modulator less than half the power of the quantum well modulator with NN operating speed = 10 GHz, and training with 45 epochs of Adagrad \cite{Duchi:EECS-2010-24} method. 
	\label{fig:nn_output_powers}}
\end{figure}

The results (Fig. \ref{fig:nn_output_powers}(a)) show that quantum well and quantum dot modulation outperforms graphene and exciton modulation in terms of accuracy in the low laser power limit. As optical power is increased, the graphene and exciton modulation approaches the accuracy of quantum well neurons. 
However, the latter performs well over a broad range of optical power inputs. However, in terms of capacitance and its effect on NN operating power, the quantum dot modulator outperforms all other modulators since it has the steepest transfer function at lowest drive voltage. (Fig. \ref{fig:nn_output_powers}(b)).


\section{Acknowledgements}
P.P., and V.S. T.E. are supported via the NSF grant 1740262 and 1740235, respectively, and SRC nCORE all under the E2CDA program. 

\end{document}